%% file: pam_arxiv.tex
\documentclass[aps,twocolumn,prb,showpacs,superscriptaddress,groupedaddress]{revtex4-1}  
\usepackage{graphicx}  
\usepackage{dcolumn}   
\usepackage{bm}        
\usepackage{amssymb}   
\usepackage{color}     
\usepackage[caption=false]{subfig}
\usepackage{amsmath}

\begin{document}



\title{Energy Scales of the  Doped Anderson Lattice Model }
\input  author_list.tex       
\date{\today}

\begin{abstract}
This paper explores the 
energy scales of the doped Anderson lattice model using dynamical
mean-field theory (DMFT), using a continuous-time Quantum Monte Carlo (CTQMC) impurity solver. 
We show that the low temperature properties of the lattice
can not be scaled using the single ion local Kondo temperature $T_K$
but instead are governed by 
a doping-dependent coherence temperature
$T^*$ which can be used to scale 
the temperature dependence of the  spectral function, transport properties, and entropy.
At half filling $T^{*}$ closely approximates the single ion
$T_{K}$, but as the filling $n_{c}$ is reduced to zero, $T^{*}$ also vanishes.
The coherence temperature $T^*$ is shown to play a role of effective impurity Kondo temperature 
in the lattice model, and physical observables show significant evolution 
at $T^*$. 
In the DMFT framework, we showed that the hybridization strength of the effective impurity model 
is qualitatively affected by the doping level, and determines $T^*$ in the lattice model. 
\end{abstract}

\maketitle


The Kondo effect was first observed as a resistivity minimum in dilute
magnetic alloys
\cite{DeHaas:1934ki}.
Jun Kondo accounted for the resistivity minimum as a consequence of an
antiferromagnetic super-exchange 
between the magnetic impurity and conduction electrons
\cite{Kondo:1964eg}. 
This antiferromagnetic coupling was later revealed to be a 
relevant coupling, renormalizing to strong coupling 
at a characteristic energy scale called the Kondo temperature 
\cite{Anderson:1970dk,Wilson:1975gl,Wilson:1974gq,Wilson:1971fb}.
Based on a strong-coupling expansion, Nozi\`eres 
showed that the ground state of a magnetically screened Kondo impurity
is described by a local Fermi liquid 
\cite{Nozieres:1974go}. 
After that, a slave-particle mean-field theory showed that Kondo
physics can be understood as the residue of a symmetry-breaking
transition that occurs in the large $N$ limit of the spin degeneracy, 
in which the Kondo temperature plays the role of a critical temperature for the phase transition
\cite{Read:1983cn,Coleman:1987gh}. 

In a large class of $f$-electron intermetallic materials called
``heavy electron'' compounds, 
such as the family of 115 compounds, 
Ce\textit{M}In$_{\textrm{5}}$ (\textit{M}=Co,Rh,Ir), 
the localized $f$ electrons form a periodic lattice of magnetic
moments whose low energy physics is described by a Kondo lattice
model \cite{Kasuya:1956daa}. 
A generic phase diagram of the Kondo lattice was proposed by Doniach 
\cite{Doniach:1977gv}, who argued that 
if the Kondo coupling is weak
the magnetic Rudermann-Kittel-Kasuya-Yosida (RKKY)
interaction overcomes the formation of Kondo singlets,
giving rise to an ordered magnetic ground-state
\cite{Ruderman:1954ug,Kasuya:1956daa,Yosida:1957gj}. 
This state has a small Fermi surface because only the conduction electrons contribute to the charge transport. 
However, if the Kondo coupling is strong, it gives 
rise to a paramagnetic ground state 
which resembles the Nozi\`eres Fermi liquid state of the Kondo impurity model. 
Such ``heavy fermi liquids'' (HFL) display carrier effective masses up to 
$\sim10^3$ times larger than in conventional metals. 
In the HFL state, the localized moments bind
to electrons, forming composite $f$-quasiparticles which hybridize with
the conduction sea, giving rise to an enlarged Fermi surface of heavy
quasiparticles. 

One of the long-standing questions concerns 
how the HFL phase evolves upon raising the temperature, and in
particular, whether additional scales, beyond the single-ion Kondo
temperature, are required to describe the gradual loss of coherence in
the HFL \cite{Settai:2007cq,Hewson:1997vc,McElfresh:1990gg,Thompson:1985fe,Stewart:1984gj}.
Theoretically, the slave-boson approach showed that an additional low energy
Fermi-liquid energy scale ($T_{FL}$) develops 
in HFL \cite{Burdin:2009fg}. 
Later numerical studies using the dynamical mean-field theory (DMFT)
confirmed that this Fermi-liquid energy scale 
exists, identifying it as the  temperature at which resistivity
develops a maximum \cite{Tanaskovic:2011jk}.
However, there is still no final 
consensus between these different studies on the precise relationship
between the coherence temperature scale and the evolution of the large
Fermi surface 
\cite{Burdin:2009fg,Tanaskovic:2011jk,Chen:2017it,Kummer:2015kq}.
These considerations motivate 
an integrated study of 
thermodynamic, transport, and spectroscopic properties of 
the Kondo lattice model, with the goal of connecting 
experimental, analytic, and numerical studies.  

In this letter, we report on a detailed study of Anderson lattice
model in the Kondo lattice regime using DMFT 
\cite{Georges:1996un,Georges:1992kt,Vollhardt:1991ht,Metzner:1989bz}, 
with a continuous-time Quantum Monte Carlo (CTQMC) impurity solver \cite{Werner:2006ko}. 
The study varied the hybridization strength, temperature, 
and the doping level to cover wide range of the phase diagram 
and investigate the scaling properties. 
Maximum-entropy methods were
used to analytically continue from imaginary to real time to 
obtain dynamical spectral functions \cite{Bryan:1990ge,Jarrell:1996is}.


The single-orbital Anderson lattice model is written as
\begin{equation}
    \begin{split}
        H
            & = \sum_{i\sigma}\epsilon_{f}f^{\dagger}_{i\sigma}f_{i\sigma} 
                + U\sum_{i}n^{f}_{i\uparrow}n^{f}_{i\downarrow} 
             - t\sum_{\langle ij \rangle \sigma}( c^{\dagger}_{i\sigma}c_{j\sigma} + H.c.) \\
            & + V\sum_{i\sigma}(c^{\dagger}_{i\sigma}f_{i\sigma} + H.c.)
             - \mu \sum_{i\sigma}(n^{f}_{i\sigma} + n^{c}_{i\sigma})
    \end{split}
\end{equation}
where $f^{\dagger}_{i\sigma}$ ($f_{i\sigma}$) is a creation (annihilation) operator 
of the $f$ electron with spin $\sigma$ at site $i$, $c^{\dagger}_{i\sigma}$ ($c_{i\sigma}$) 
is a creation (annihilation) operator of the conduction electron with spin $\sigma$ at site $i$, 
and $n^{\alpha}_{i\sigma}=\alpha^{\dagger}_{i\sigma}\alpha_{i\sigma}$ ($\alpha=f,c$).

For convenience, all energy scales are written in 
units of $D$, 
the half bandwidth of the conduction band, and the Boltzmann constant
$k_{B}$ is set to unity. 
We considered a two-dimensional square lattice with half bandwidth
$D=4t$. 
To achieve the Kondo lattice regime, we place the f-level at the
bottom of the band, choosing $\epsilon_{f} = -1.0$ and se
$U=2.0$, so that the energy of the doubly occupied state
$\epsilon_{f}+U=1.0$ lies at the top of band. 
The hybridization $V$, chemical potential $\mu$, and inverse
temperature $\beta$ were varied from
$0.18$ to $0.54$, $-0.8$ to $0.8$, and $80.00$ to $200.00$, respectively.


\begin{figure}[htbp]
    \centering
    \subfloat[]{\includegraphics[width=.24\textwidth]{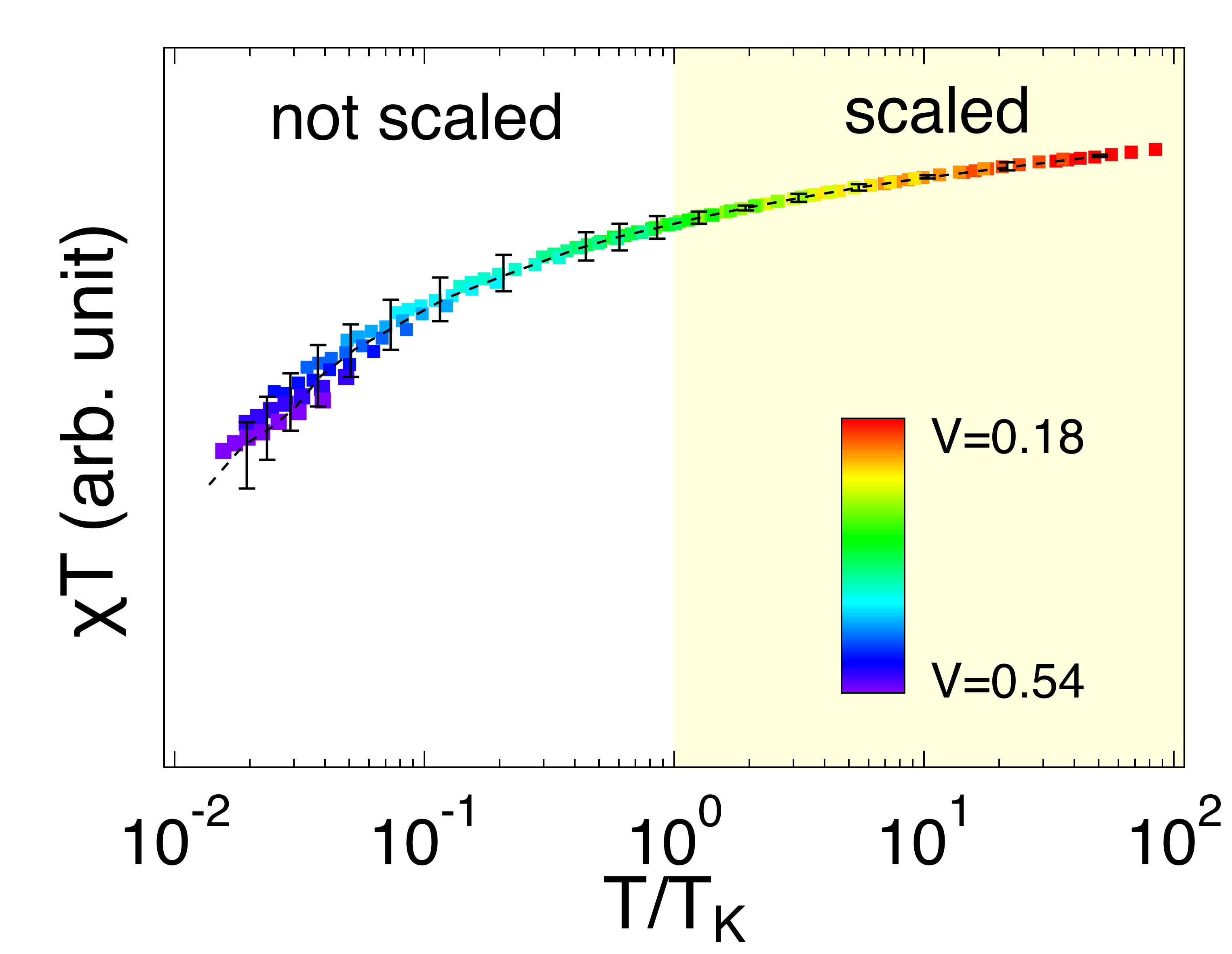}}
    \subfloat[]{\includegraphics[width=.24\textwidth]{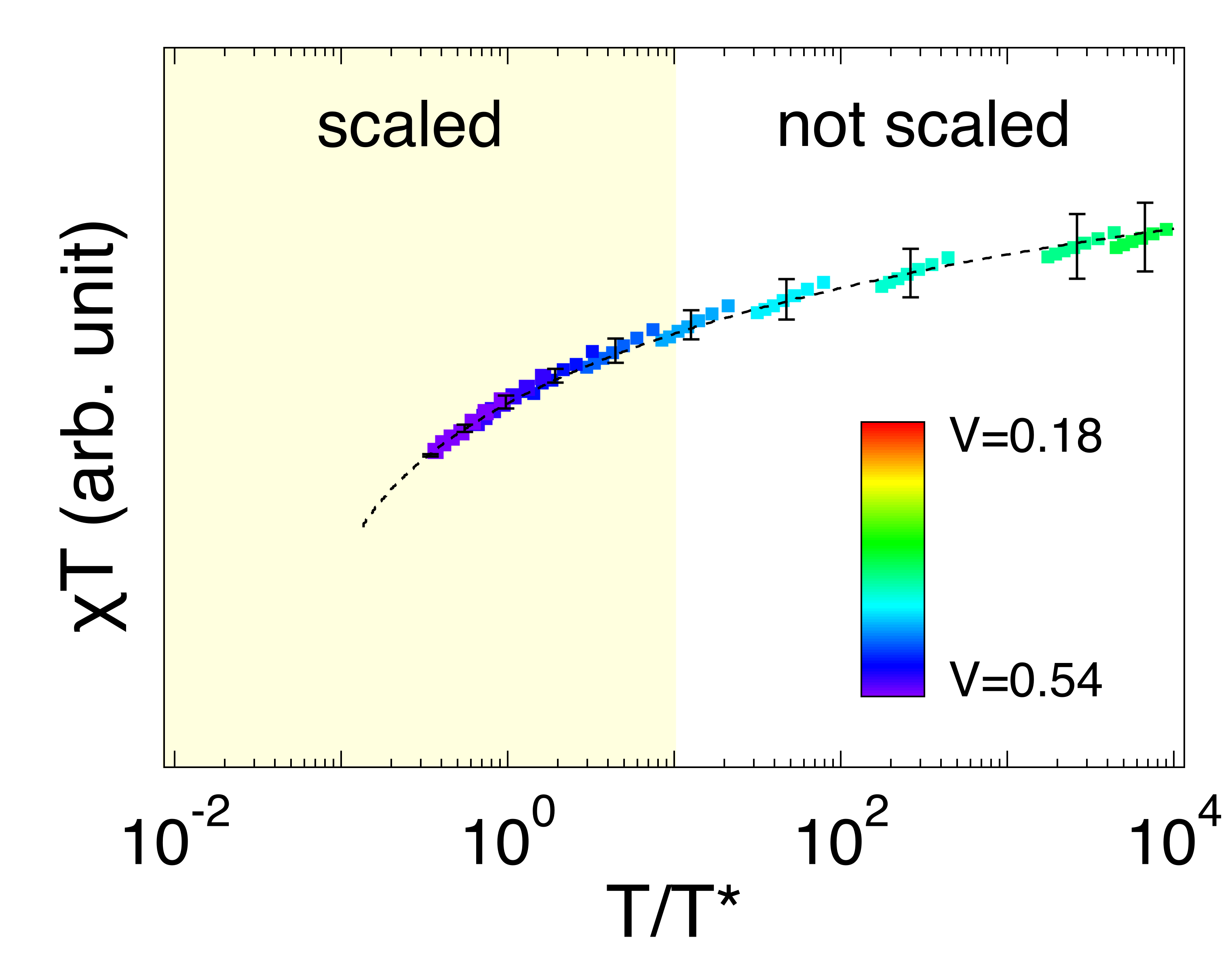}}
    \newline
    \noindent\subfloat[]{\includegraphics[width=.49\textwidth]{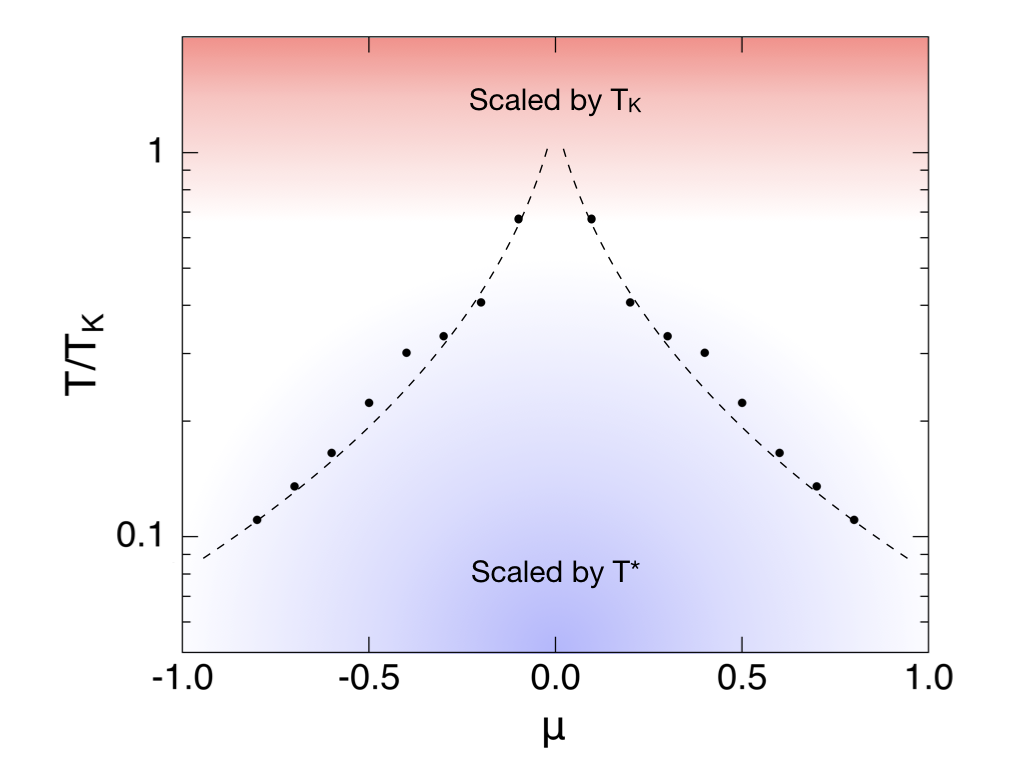}}
    \caption{Local static spin susceptibility of the Anderson lattice
    model scaled 
(a) by $T_K$ 
        and (b) by $T^*$, computed at 
        chemical potential $\mu=-0.8$, for a range of 
hybridization between $V=0.18$ (red) and $V=0.54$ (purple).
        The dashed lines are the best-fit and the error bars show the mismatch 
        between the best-fit line and actual data.
(c)
        Schematic phase diagram showing the variation of $T^{*}$
        with chemical potential, and the regions where the data scales
	with $T_{K}$ (red) and with $T^{*}$ (blue). The 
        dashed line is a guide to the eye. 
        }
    \label{fig: definition}
\end{figure}

Figure 1 (a) shows the 
local spin susceptibility $\chi_{loc}(\omega=0)$ for
$\mu=-0.8$, for a variety of hybridization values $V$, scaled by 
the single-impurity Kondo temperature $T_K$ evaluated 
with the same parameters, defined by
\begin{equation}
    T_K = \sqrt{2J_K \rho }\exp\left[-\frac{1}{2J_{K}\rho } \right]
\end{equation}
where $\rho $ is the density of states per spin 
of the conduction band at the Fermi level and 
$J_K = (|\epsilon_{f}-\mu|^{-1} + |\epsilon_{f}-\mu+U|^{-1})V^2$ is 
the Kondo exchange \cite{Schrieffer:1966hu}. 
Because the undoped model ($\mu=0.0$) is particle-hole symmetric,  
electron and hole doped cases behave identically
\cite{supp}.
The scaling collapse of the susceptibility curves at high temperatures
$\chi_{loc} (T)\sim \frac{1}{T}f (T/T_{K})$
shows that the high temperature physics of the 
Anderson lattice model is scaled by the single-ion Kondo temperature,
regardless of the doping level \cite{supp}, implying that
the high temperature physics at $T> T_{K}$ is that of a single 
impurity model.

However the local susceptibility (Fig. 1 (a)) does not
scale with the single-ion Kondo temperature at low temperatures.
To scale the low-$T$ regime, we define a 
coherence temperature  $T^*$, parameterized as
\begin{equation}
  T^{*} = \sqrt{2J^{latt} \rho }\exp\left[-\frac{1}{2J^{latt}\rho } \right]
\end{equation}
where $J^{\text{latt}}=jJ_K$ is an effective Kondo 
lattice exchange strength.
The unique fitting parameter $j$ is adjusted at each doping level to 
collapse the low temperature susceptibilities  
onto a single curve
\cite{supp}.
Figure 1 (b) shows that the low-$T$ susceptibilities are successfully scaled by $T^*$ 
with $j= 0.3$.
The emergence of the temperature scale $T^*$ indicates that the Kondo lattice model behaves 
differently in the fundamental level at low-$T$ regime. 


Figure 1 (c) shows how $T^*$ varies as the chemical potential is changed. 
When $n_c$ is close to $0$, $T^*$ is suppressed towards zero while
when $n_c$ is close to half-filling, $T^*$ tends towards the
single-ion Kondo temperature $T_{K}$, a result that agrees with
previous studies \cite{Tanaskovic:2011jk}.

\begin{figure}[htbp]
    \centering
    \subfloat[]{\includegraphics[width=0.99\columnwidth]{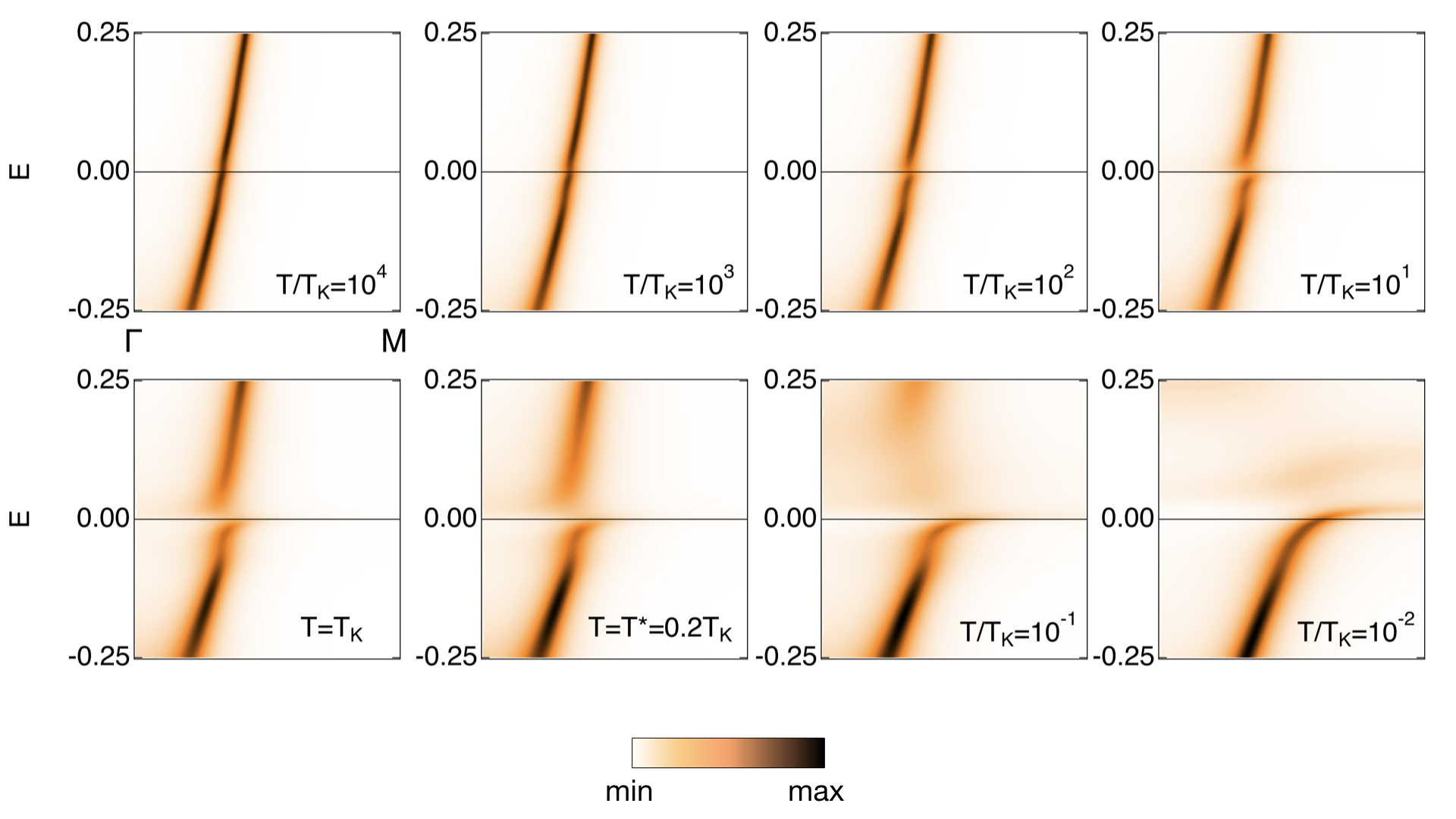}}
    \newline
    \noindent
    \subfloat[]{\includegraphics[width=0.99\columnwidth]{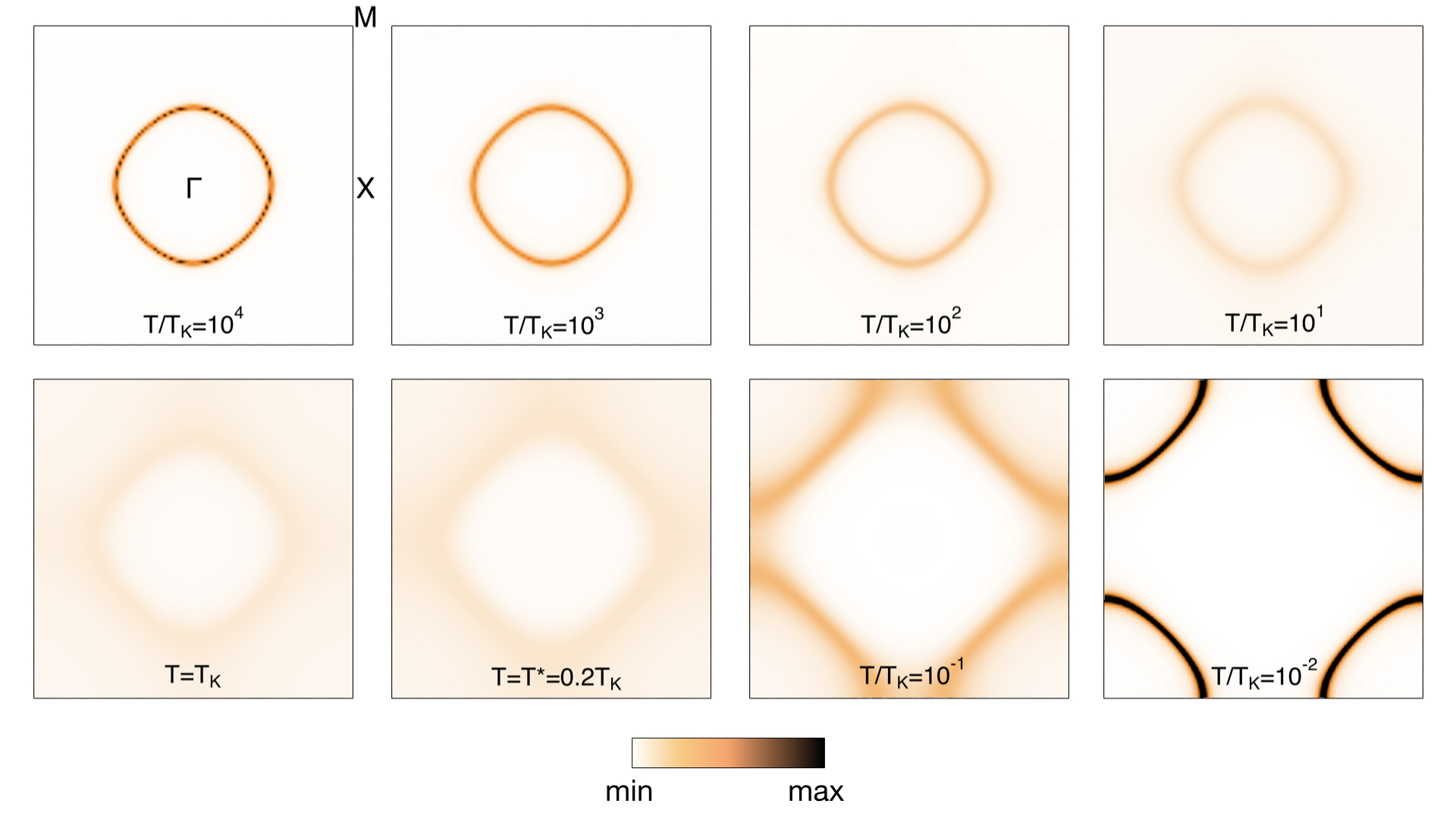}}
    \caption{Intensity plots showing the momentum and energy resolved 
    conduction electron spectral
    function (a) as a function of momentum and (b) at fixed $E=0$, 
    showing the evolution of the Fermi surface with temperature, for
    $\mu=-0.5$. 
    }
    \label{ label}
\end{figure}

Figure 2 (a) shows the calculated momentum- and energy-resolved
total spectral function 
\begin{equation}
    A(\vec{k}, \omega) = \frac{1}{\pi} {\rm Im} \Big[G_{f}(\vec{k}, \omega-i\delta ) + G_{c}(\vec{k}, \omega-i\delta )\Big]
\end{equation}
at $\mu=-0.5$ case.  At high temperatures, only the
coherent conduction band is observed near the Fermi level.  Lowering the
temperature, an incoherent $f$-electron spectrum develops at
the Fermi level as a sign of Kondo singlet formation.  It is
notable that the spectral function starts to change far above
the local Kondo temperature.  It agrees well with recent ARPES
measurement on the Ce-115 heavy fermion compound
\cite{Jang:2017uv,Chen:2017it}.  Crossing through $T_K$, the spectra near the
Fermi level becomes incoherent, and the velocity of the
ill-defined quasiparticles gets smaller as the $f$-electron develops
near the Fermi energy. The spectrum is maximally incoherent at $T=T^*$, and the
quasiparticle band only re-establishes its coherence 
below $T^*$.

Figure 2 (b) shows the evolution of the Fermi surface. 
Starting from a coherent small Fermi surface at high temperatures, 
it continuously evolves into an incoherent large Fermi surface, which sharpens
well below the coherence temperature $T^*$. 
This continuous, but non-monotonic evolution of the Fermi surface
gives a hint for nature of the 
non-Fermi liquid phase observed in the quantum critical region.

\begin{figure}[htbp]
    \centering
    \subfloat[]{\includegraphics[width=.24\textwidth]{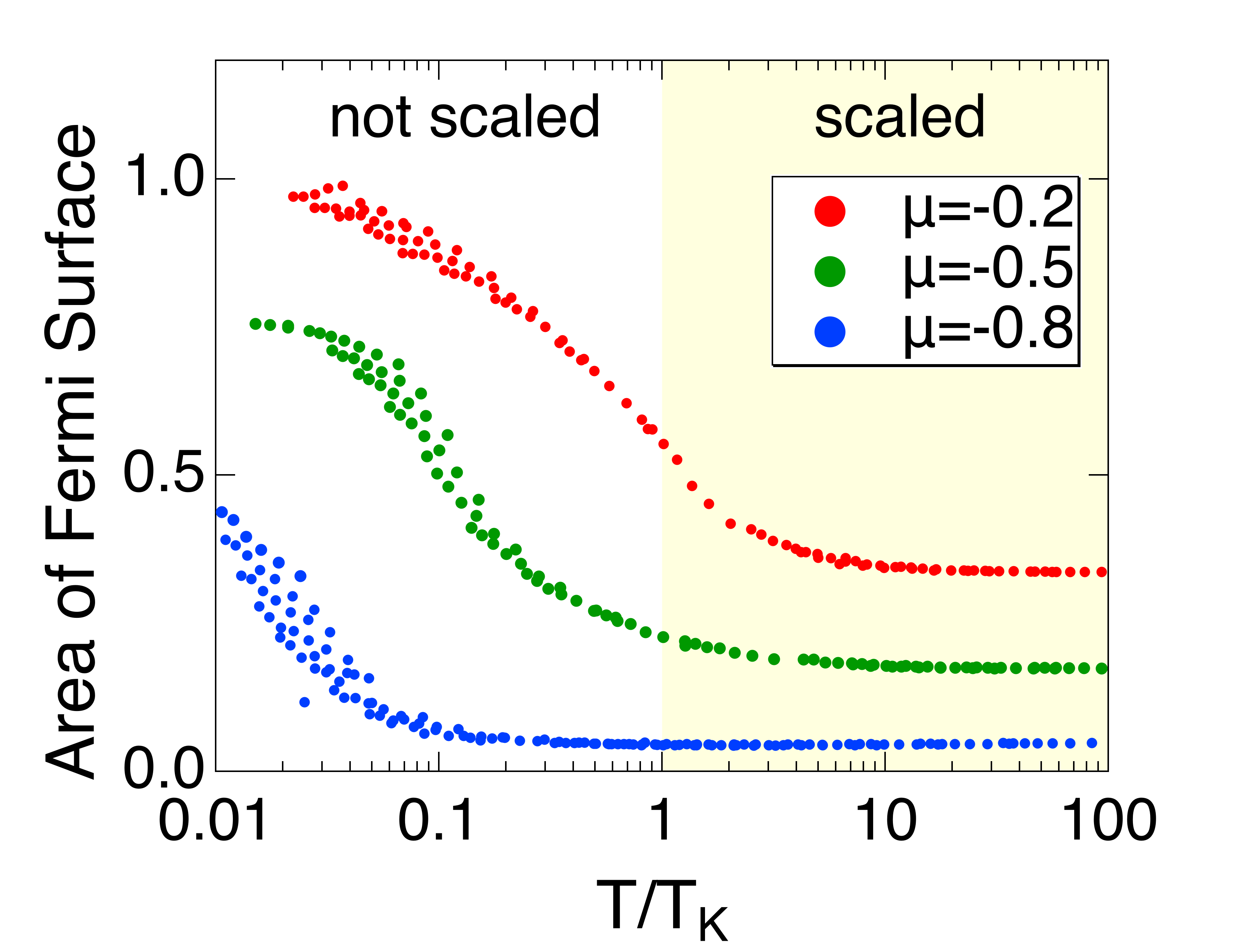}}
    \subfloat[]{\includegraphics[width=.24\textwidth]{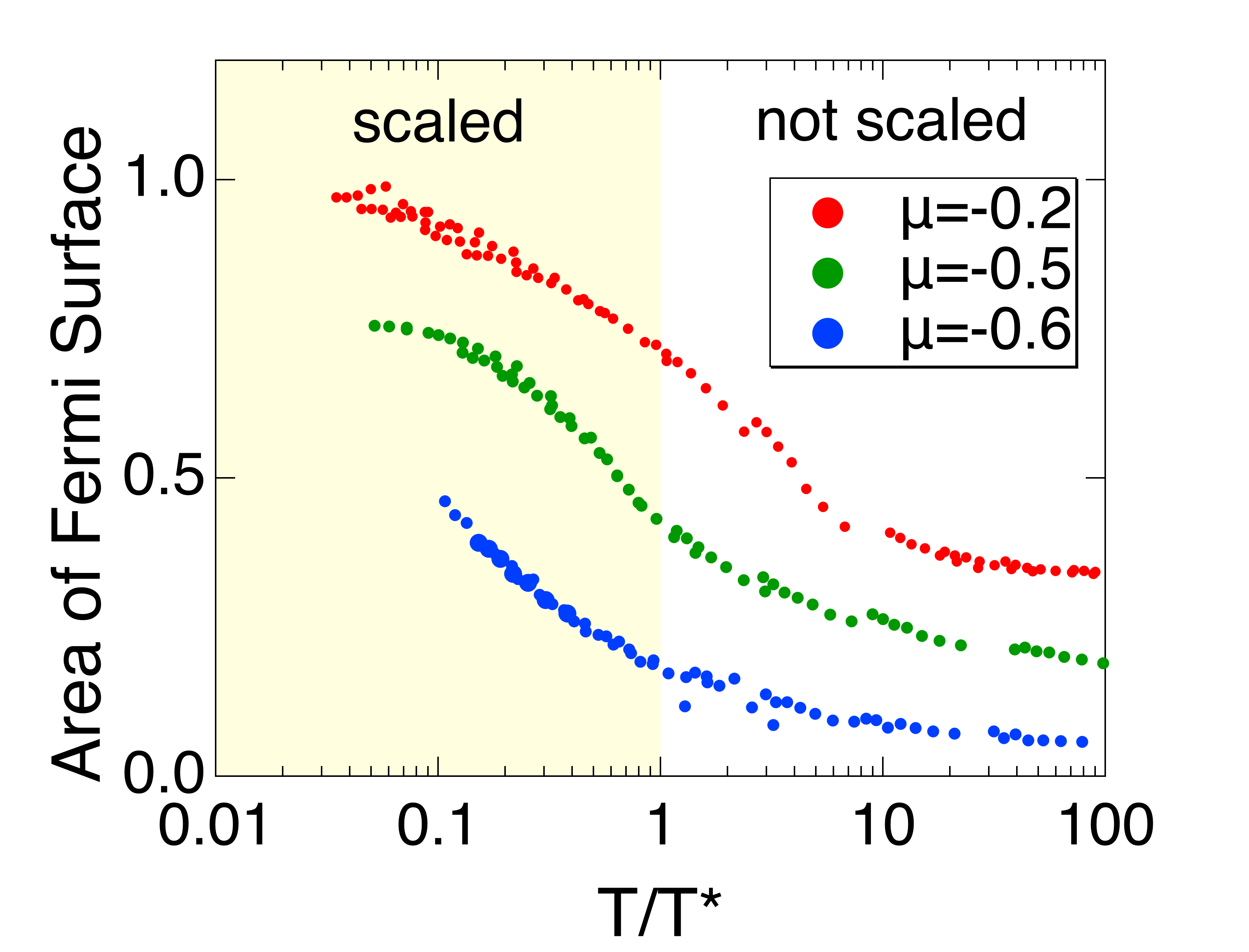}}
    \newline\noindent
    \subfloat[]{\includegraphics[width=.24\textwidth]{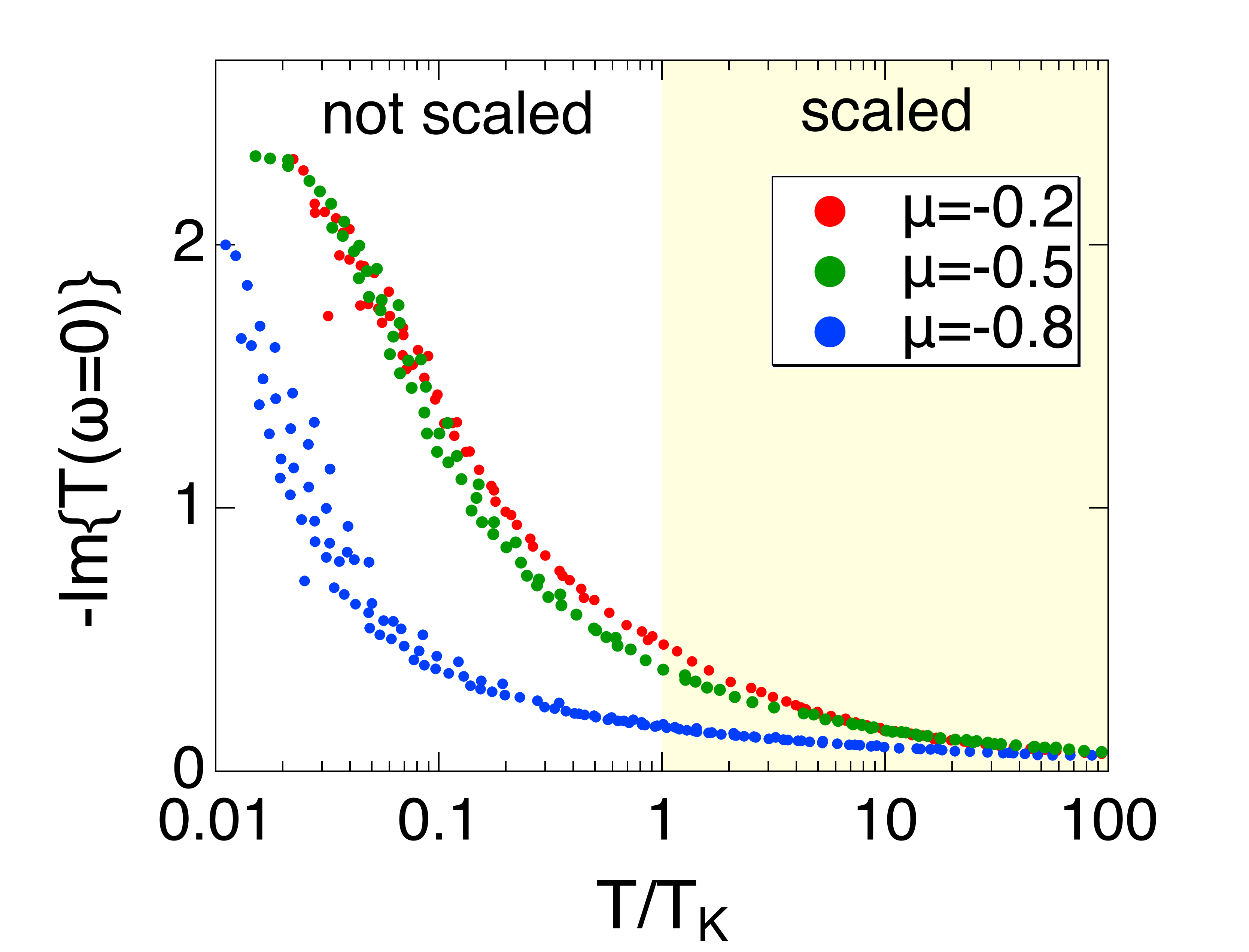}}
    \subfloat[]{\includegraphics[width=.24\textwidth]{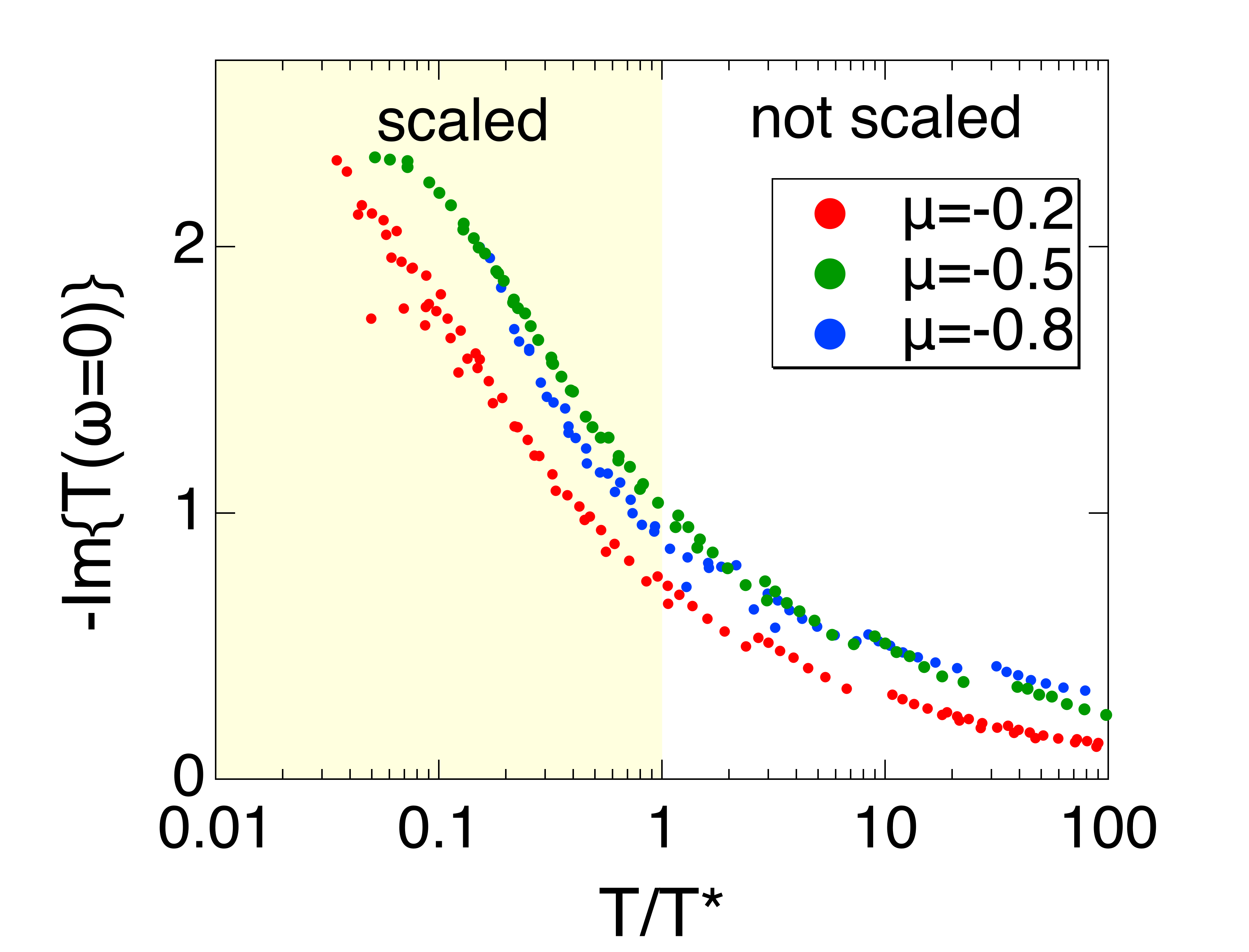}}
    \caption{Area of the Fermi surface (a,b) and imaginary part of the \textit{T}-matrix of the conduction
        electrons at the Fermi level (c,d) scaled by $T_K$ and $T^*$.
        }
    \label{ label}
\end{figure}

Figure 3 shows the area of the Fermi surface (a,b) and imaginary part of the \textit{T}-matrix ($T=V^2G_f$) 
of the conduction band at the Fermi level (c,d) 
scaled by $T_K$ and $T^*$. 
Even though $T_K$ scales the high-$T$ behavior of both observables, 
there is no significant feature in both observables at $T=T_K$. 
For example, the small Fermi surface of the $\mu=-0.8$ case does not evolve 
to the large Fermi surface phase until far below $T=T_K$. 
In contrast, both Fermi surface area and the $f$-electron DOS at the
Fermi level evolve rapidly around 
$T=T^*$, regardless of the chemical potential. 



\begin{figure}[htbp]\label{fig:rho}
    \centering
    \subfloat[]{\includegraphics[width=.24\textwidth]{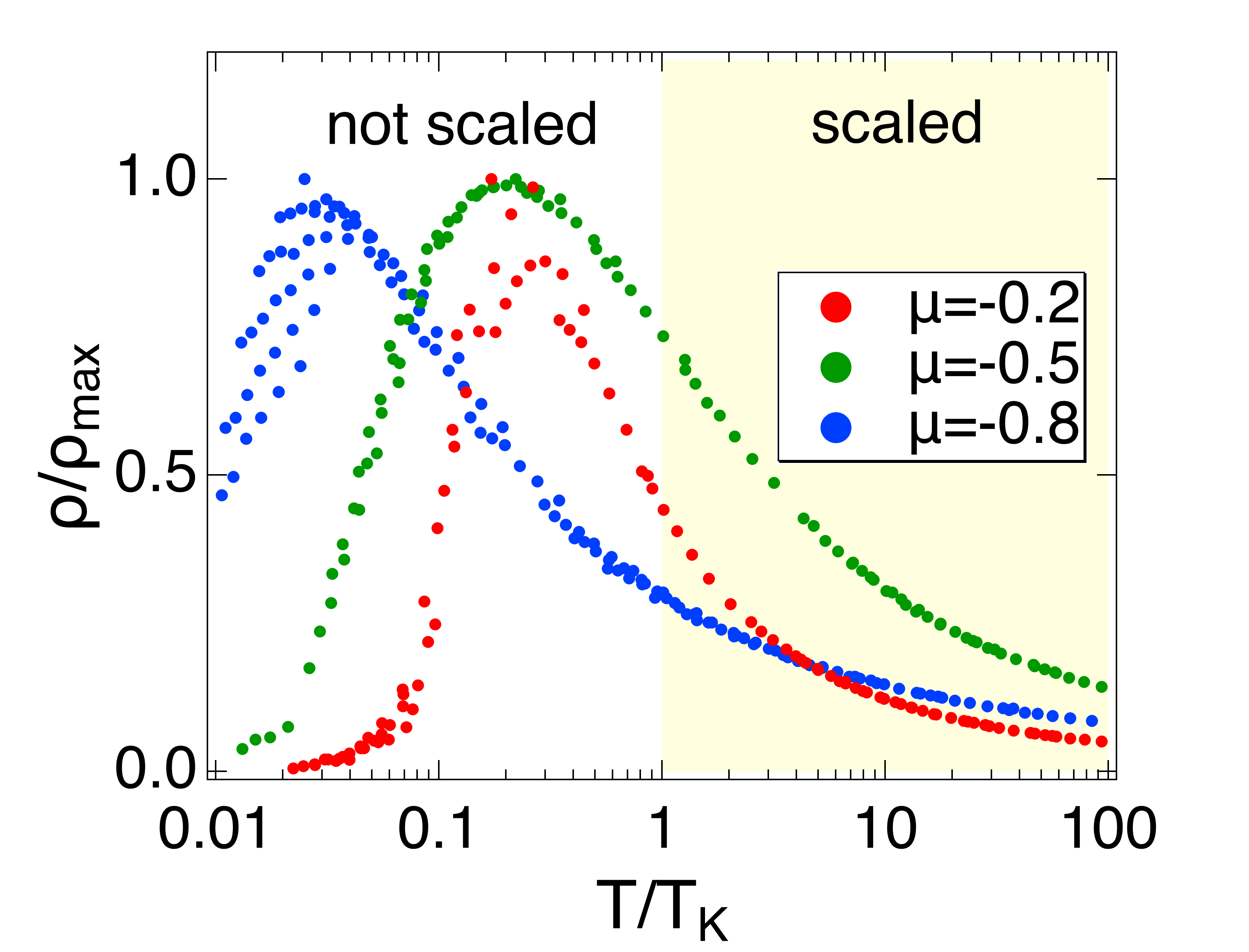}}
    \subfloat[]{\includegraphics[width=.24\textwidth]{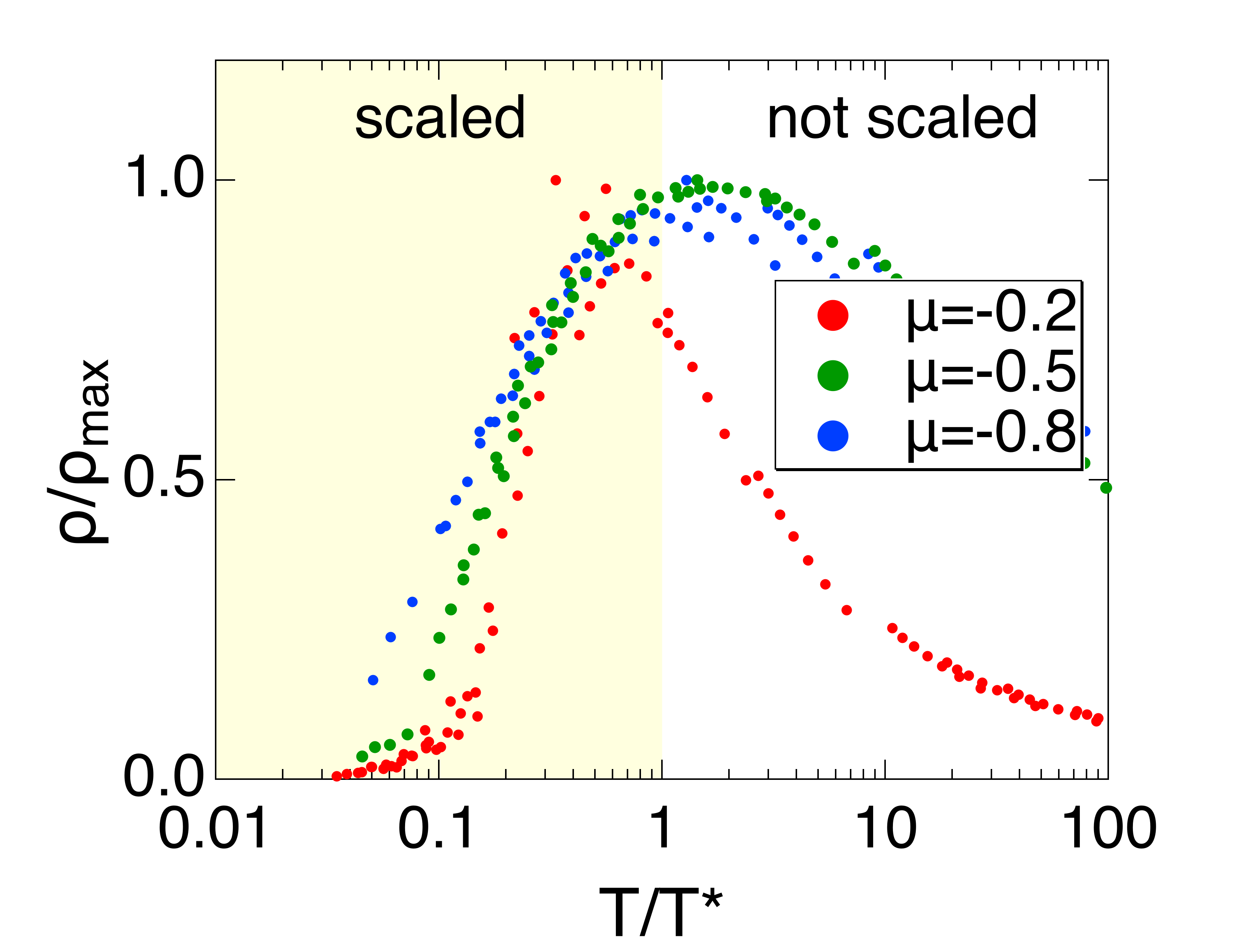}}
    \newline
    \subfloat[]{\includegraphics[width=.24\textwidth]{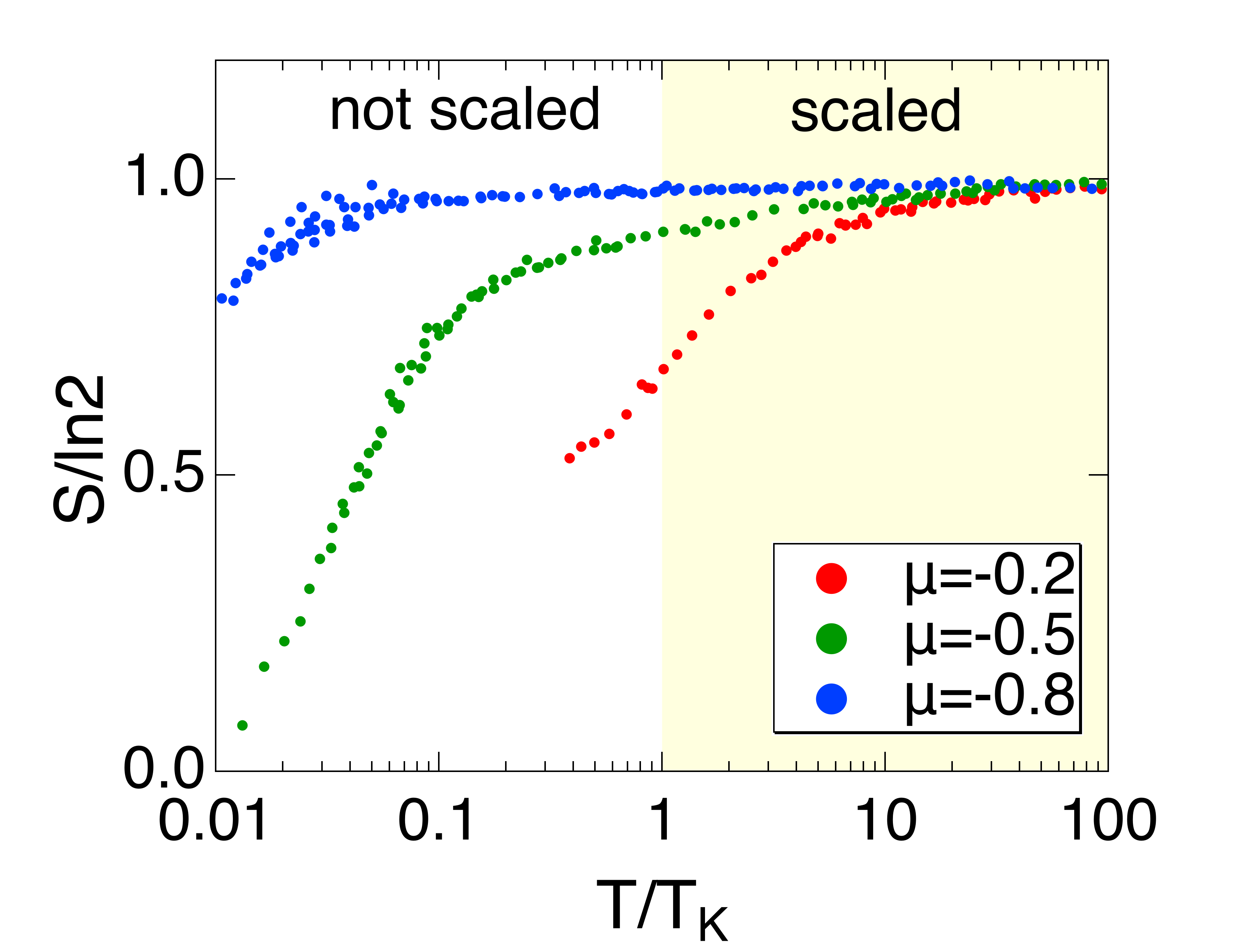}}
    \subfloat[]{\includegraphics[width=.24\textwidth]{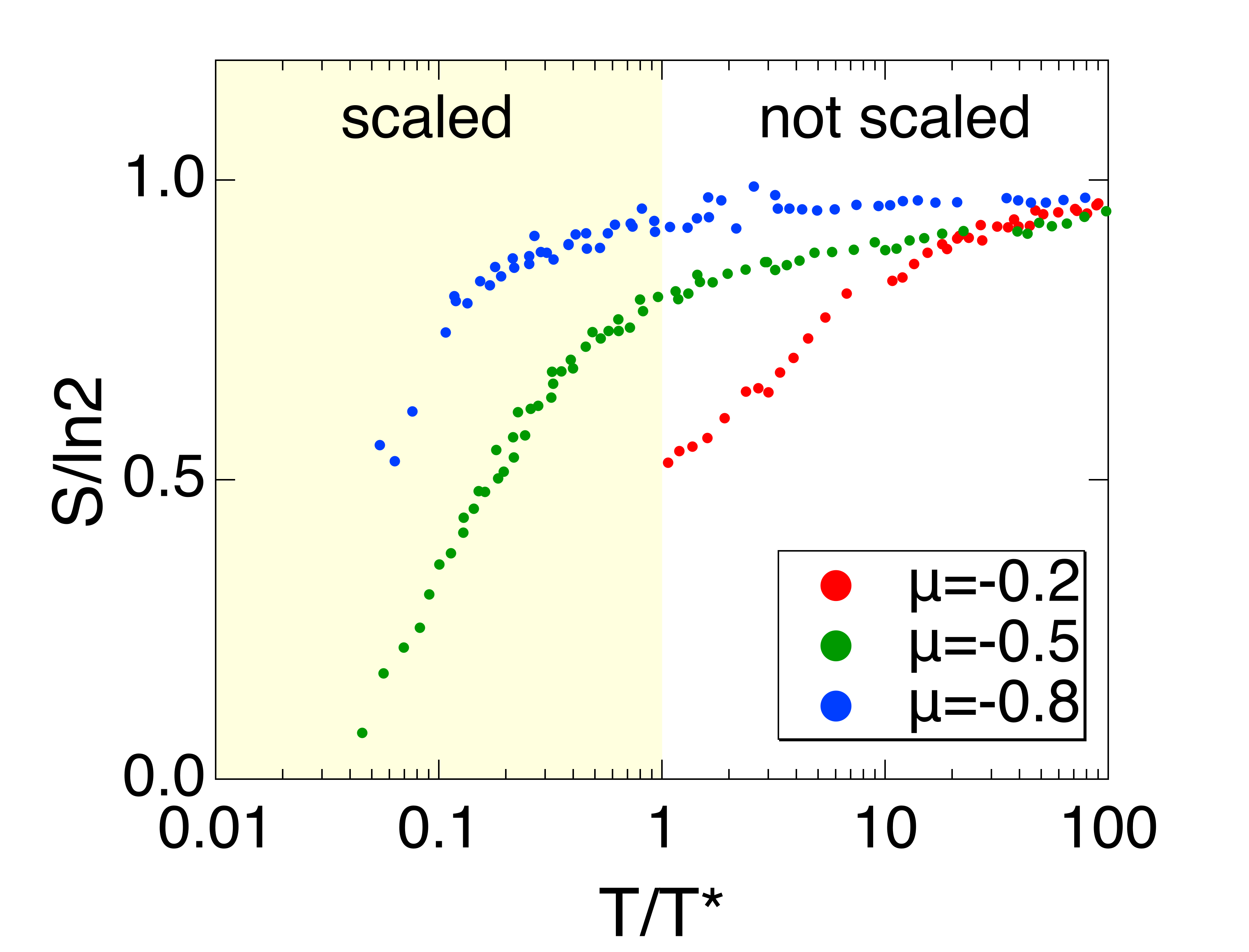}}
    \caption{Resistivity (a,b) and entropy (c,d) scaled by $T_K$ and $T^*$.
            }
    \label{ label}
\end{figure}
The coherence temperature $T^*$ also plays a significant role in the
transport properties.  Figure 4 shows the resistivity of
$\mu=-0.2,-0.5,-0.8$ cases.  In the high-$T$
regime, the temperature dependence of the resistivity at different
hybridization strengths can be scaled with the local Kondo temperature
$T_K$ as Fig. 4 (a).
As the temperature is reduced, the resistivity reaches a maximum and
decreases forming a coherent HFL state. 
Figure 4 (b) shows that the low-$T$ resistivity is scaled
by the coherence temperature $T^*$.  In addition, the calculated
resistivity develops its maximum value at temperatures $T\sim T^*$
which lie below the single ion 
$T_K$.  This suggests that experimentally observed resistivity maxima
are related to $T^*$ and can be used to estimate this quantity. 

To investigate the screening of the local moments more directly,
we also calculated the entropy of the impurity degree of freedom $S$.
In Fig. 4 (c,d), the high temperature
entropy approaches $\ln2$ per site, corresponding to the unscreened local 
moments of the $f$-electrons.  It is remarkable that the entropy
remains of order $\sim\ln2$ even at $T<T_K$ in the heavily doped case
($\mu=-0.8$) in Fig. 4 (c), indicating that the local
moments  are largely unscreened around $T=T_K$.  Instead,
as shown in Fig. 4 (d), the entropy starts to drop around
$T=T^*$ regardless of the doping level, although the amount of suppressed entropy
depends on the doping level. 
The difference in the amount of suppressed entropy derives from 
the conduction electron occupancy $n_c$. 
Previous studies of the strong-coupling limit 
of the Kondo lattice model suggest that an entropy of order $n_c
\ln(2)$ is lost on passing through the Kondo temperature $T_K$ \cite{Lacroix:1985jz}.
However, our results show that the suppression of magnetic entropy 
$S_M \sim n_c \ln(2)$ occurs at temperatures around $T^{*}$, rather
than $T_{K}$. 
$T^{*}$ thus sets the 
characteristic scale at which the moments become entangled with the 
conduction sea in the lattice.

In the DMFT framework, the Anderson lattice is treated as 
an effective impurity embedded in cavity with a self-consistently determined 
conduction electron bath.
Figure 5 shows the self-consistent hybridization strength 
$\Delta^{\text{latt.}}_0 \equiv \rm Im \Delta^{\text{eff.}}(z=0)$,
normalized by the bare hybridization strength of the model
$\Delta_0\equiv \rm Im \Delta(z=0)$. 
\begin{figure}[htbp]
    \centering
    \includegraphics[width=.49\textwidth]{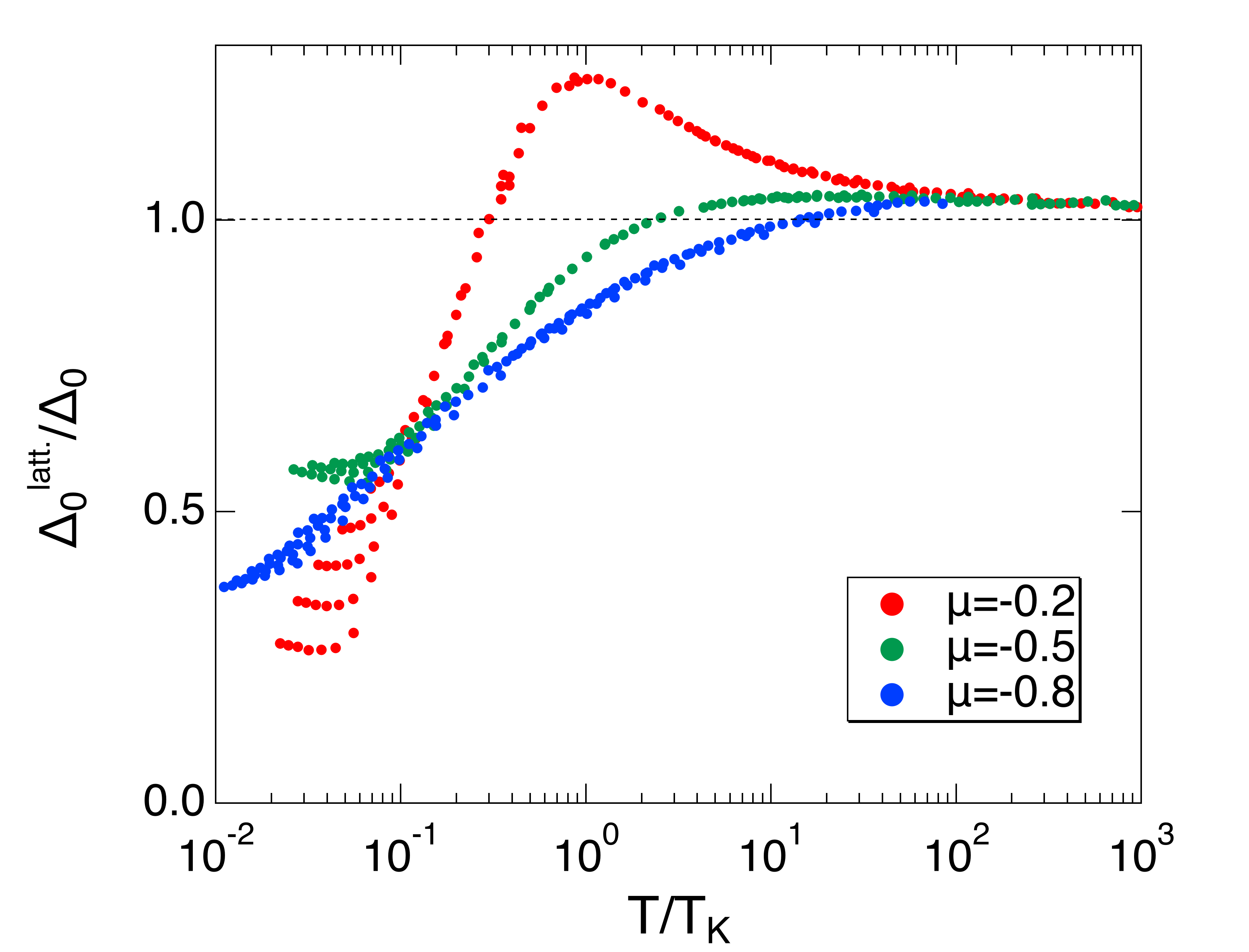}
    \caption{Ratio between effective 
            hybridization strength of the lattice model ($\Delta^{\text{latt}}_0$) and 
            hybridization strength of the impurity model.
            }
    \label{ label}
\end{figure}
In the lightly doped cases, ($\mu=-0.2$)
the effective hybridization function is enhanced at intermediate
temperatures, but regardless of doping, as the 
temperature decreases, the effective 
hybridization strength is significantly suppressed.
This is due to the formation of a pseudo-gap structure in the cavity
electronic density of states. 
As the pseudo-gap structure arises, the bath electron density of states 
at the Fermi level $\rho$ decreases. 
This reduces the coupling constant $J_K \rho$ and the effective lattice 
Kondo coupling $J^{latt}$ which determines the coherence scale 
\begin{equation}
  T^{*} = \sqrt{2J^{latt} \rho }\exp\left[-\frac{1}{2J^{latt}\rho } \right],
\end{equation}
becomes smaller as a result.



In conclusion, we have studied the temperature scales of the doped
Anderson lattice model using single-site 
dynamical mean field theory.  The local Kondo temperature $T_K$ defined
by the Kondo exchange coupling $J_K$ governs the high-temperature regime, but
a new scale  $T^*$,  defined by a modified Kondo lattice
exchange coupling $J^{\text{latt.}}$ governs the low-$T$ regime.
$T^*$ has clear doping dependency, and it approaches to zero as $n_c$
goes to zero, but tends to the single-ion $T_K$ as $n_c$ approaches 
half filling.
Various physical observables such as spectral function and transport
properties are scaled by $T_K$ at high-$T$ regime, and $T^*$ at
low-$T$ regime.

We have also confirmed that most observables show a significant
change at $T^*$, which is always significantly smaller than $T_K$. 
The DMFT self consistency determines
the suppression and magnitude of $T^*$.

This research was supported by Basic Science Research Program through 
the National Research Foundation of Korea(NRF) funded by the Ministry
of Education (2017R1D1A1B03032069), and by DOE Basic energy sciences grant
DE-FG02-99ER45790 (PC). 



\end{document}




\title{Supplementary Information: Energy Scales of Doped Anderson Lattice Model }
\input  author_list.tex       
\date{\today}

\maketitle

\begin{figure}[htbp]
    \centering
    \subfloat[]{\includegraphics[width=.24\textwidth]{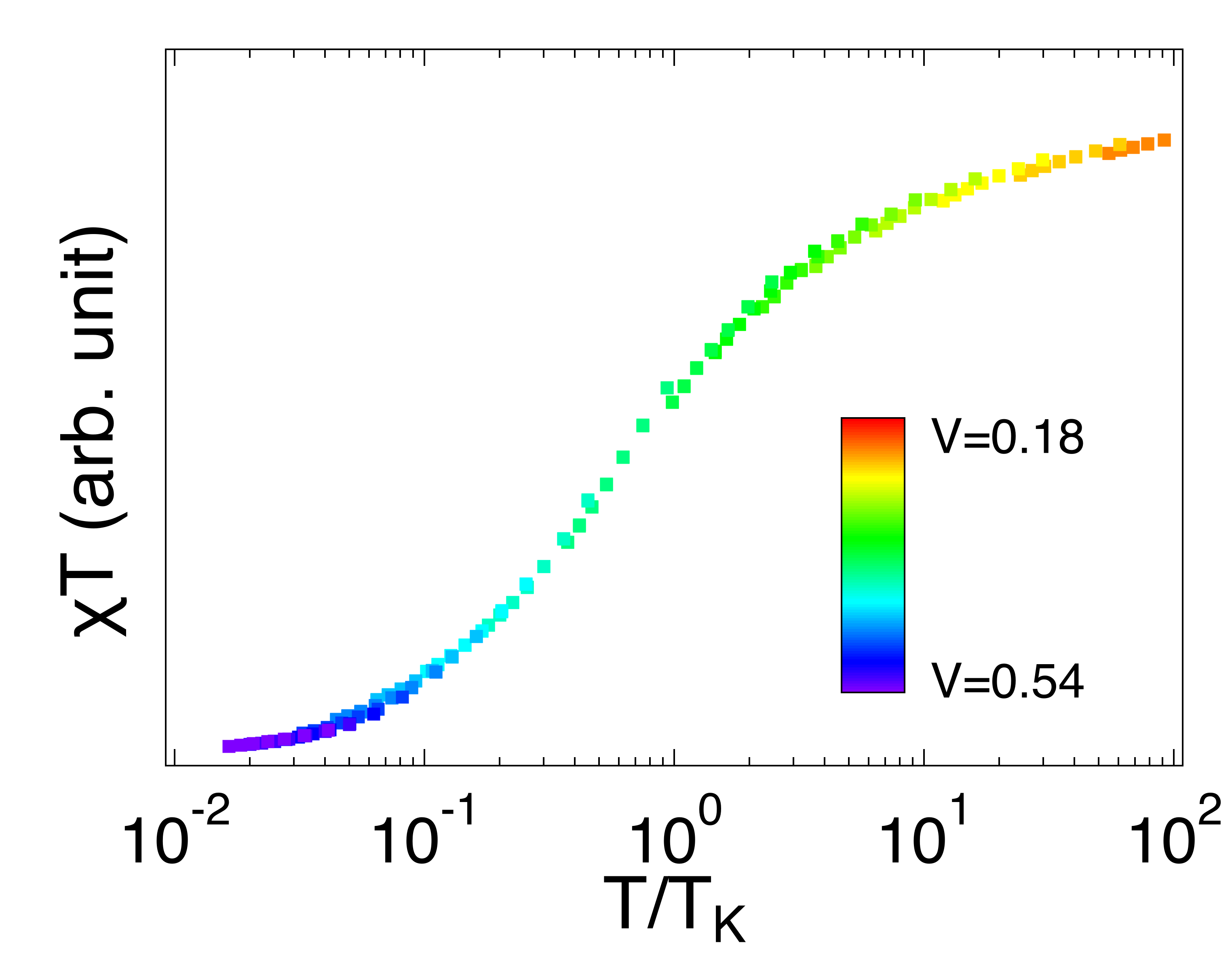}}
    \noindent\subfloat[]{\includegraphics[width=.24\textwidth]{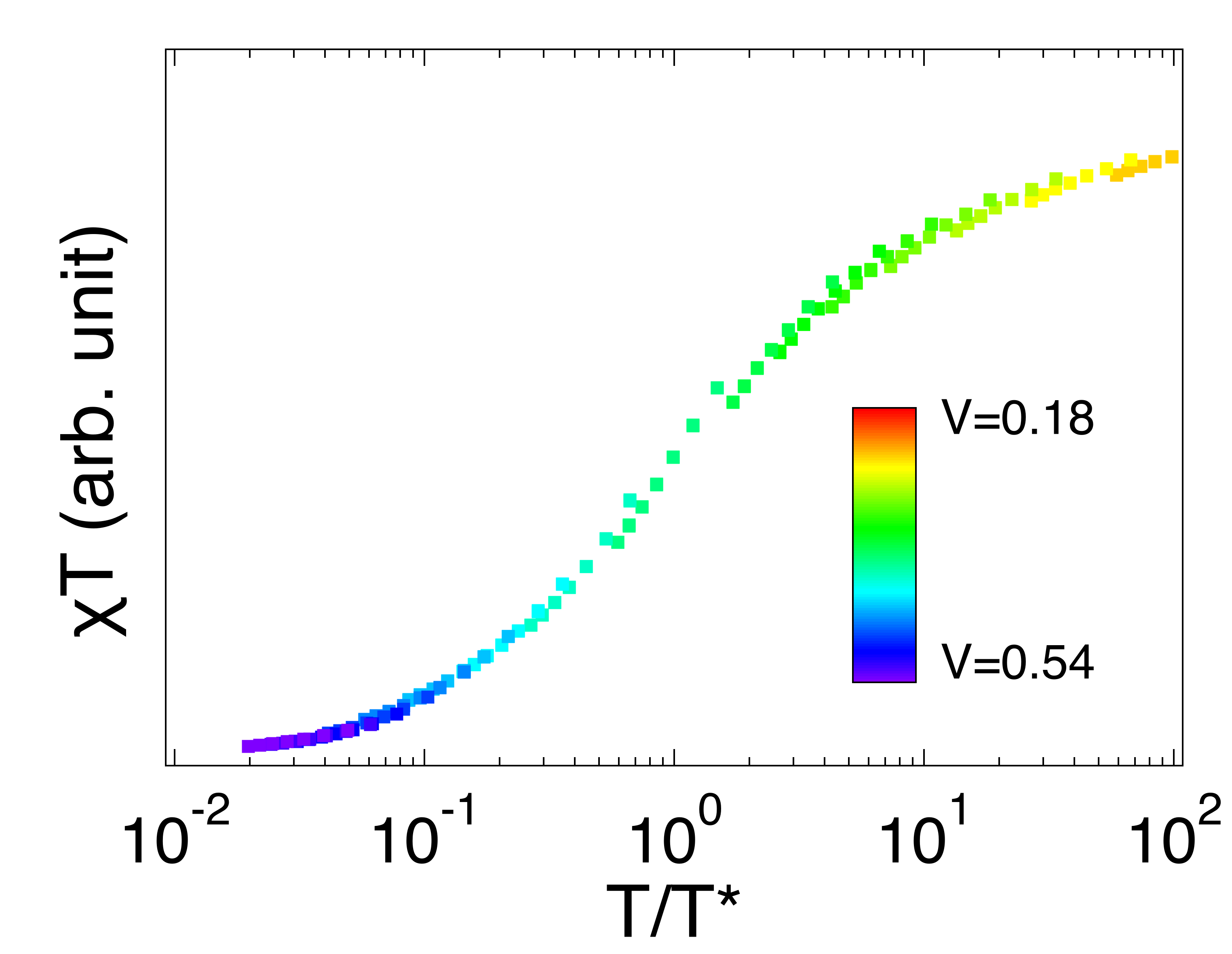}}
    \newline
    \subfloat[]{\includegraphics[width=.24\textwidth]{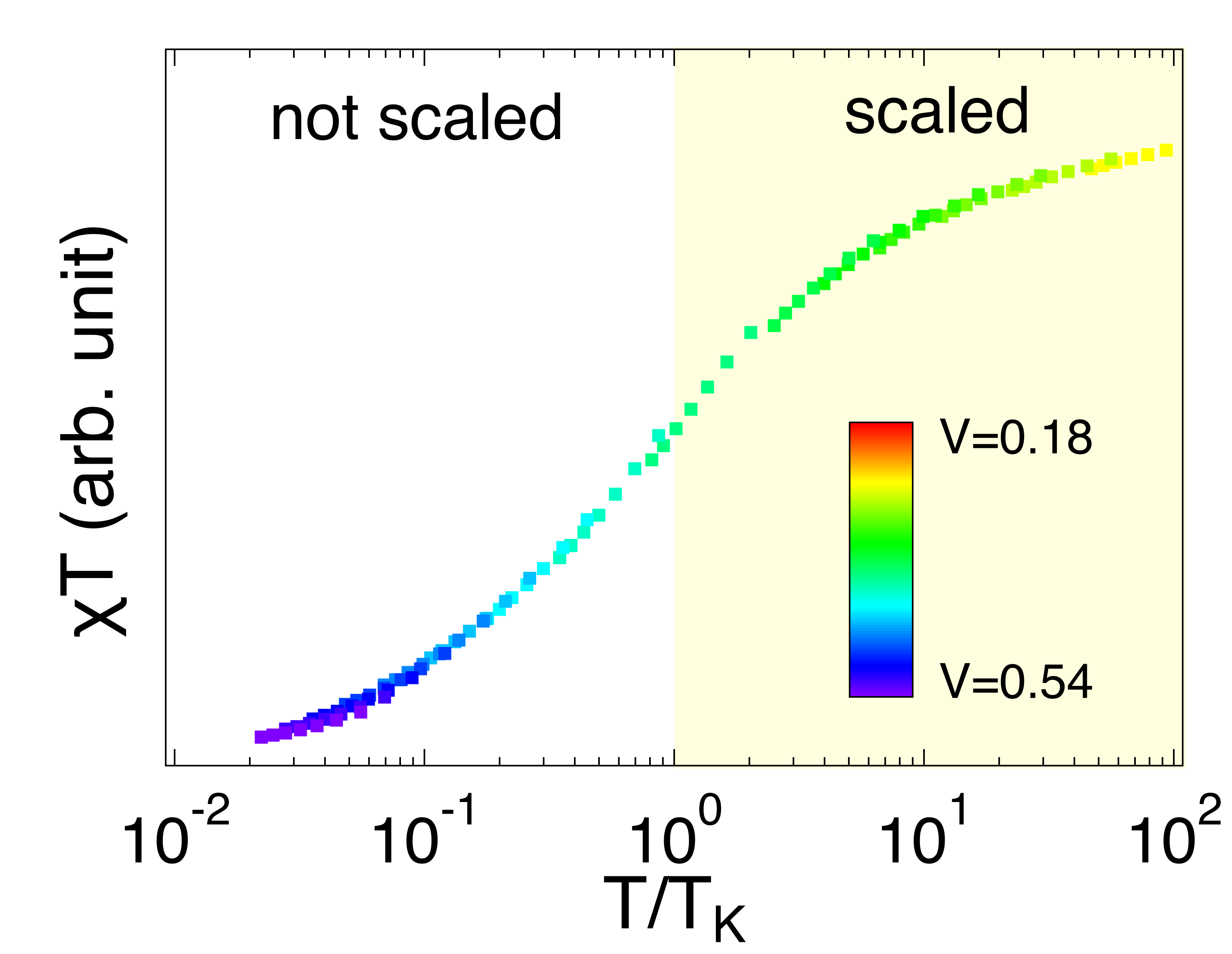}}
    \noindent\subfloat[]{\includegraphics[width=.24\textwidth]{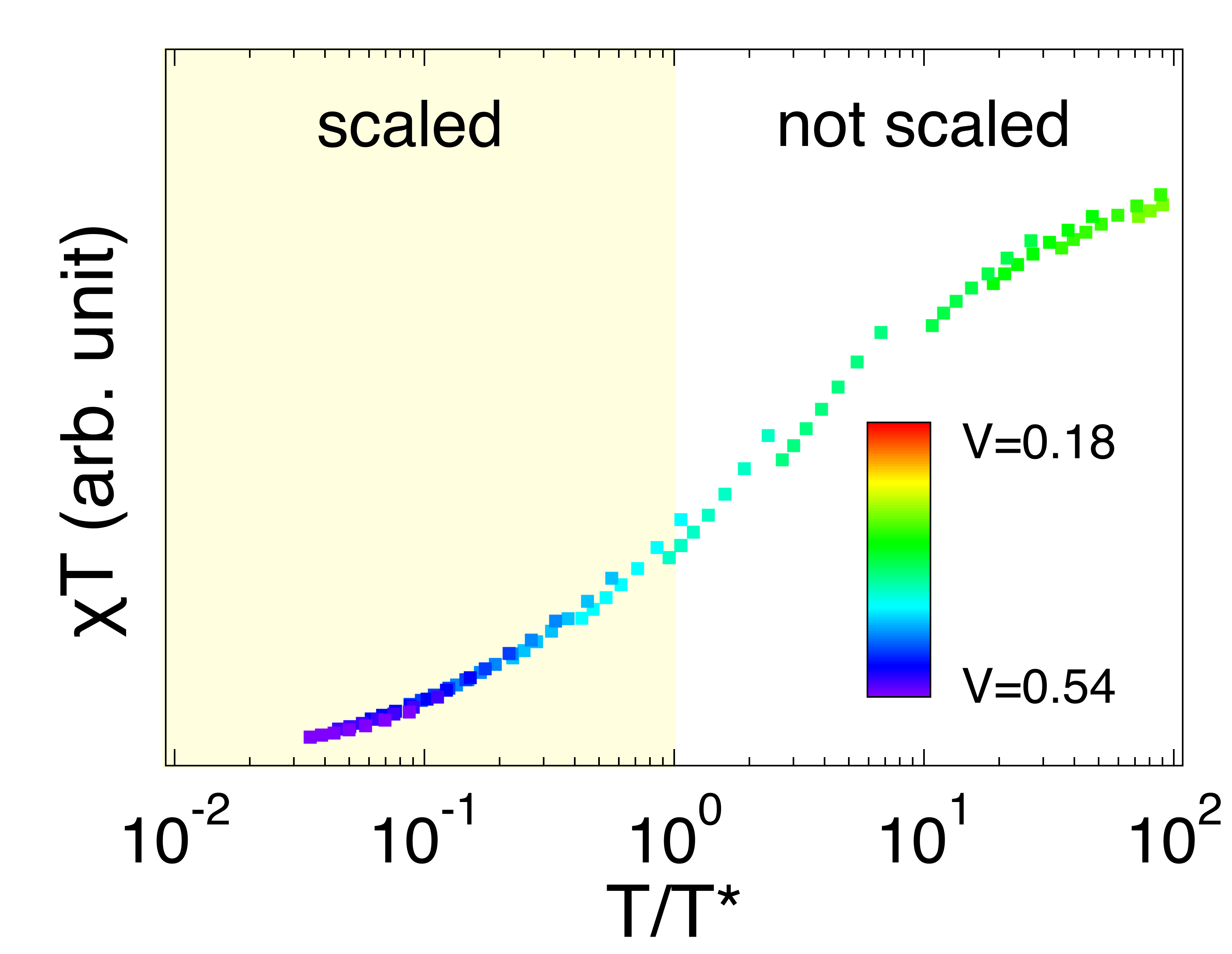}}
    \newline
    \subfloat[]{\includegraphics[width=.24\textwidth]{supp_chis0_mu-05_tk.png}}
    \noindent\subfloat[]{\includegraphics[width=.24\textwidth]{supp_chis0_mu-05_tcoh.png}}
    \caption{Local static spin susceptibility scaled by $T_K$ (a,c,e) 
        and $T^*$ (b,d,f) of the Anderson lattice model with
        chemical potential of $\mu=-0.1,-0.2,-0.5$ (from top to bottom)
        }
\end{figure}

In addition to the $\mu=-0.8$ case in the main text, the local Kondo temperature $T_K$ scales 
the high-$T$ $\chi_{loc}(\omega=0)$ in all chemical potentials as shown in Fig. 1.
Figure 1 (a-f) show the high-$T$ and low-$T$ scaling behavior of $\chi_{loc}(\omega=0)$ with $\mu=-0.1,-0.2,-0.5$.
The low-$T$ scaling is achieved by defining the coherence temperature scale $T^*$ as mentioned in the main text. 
In the case of $\mu=-0.1$, local Kondo temperature $T_K$ and coherence temperature $T^*$ are virtually 
indistinguishable as $T_K$ seemingly scales both low- and high-$T$ regime as shown in Fig. 1 (a) and (b).

\begin{figure}[htbp]
    \centering
    \subfloat[]{\includegraphics[width=.24\textwidth]{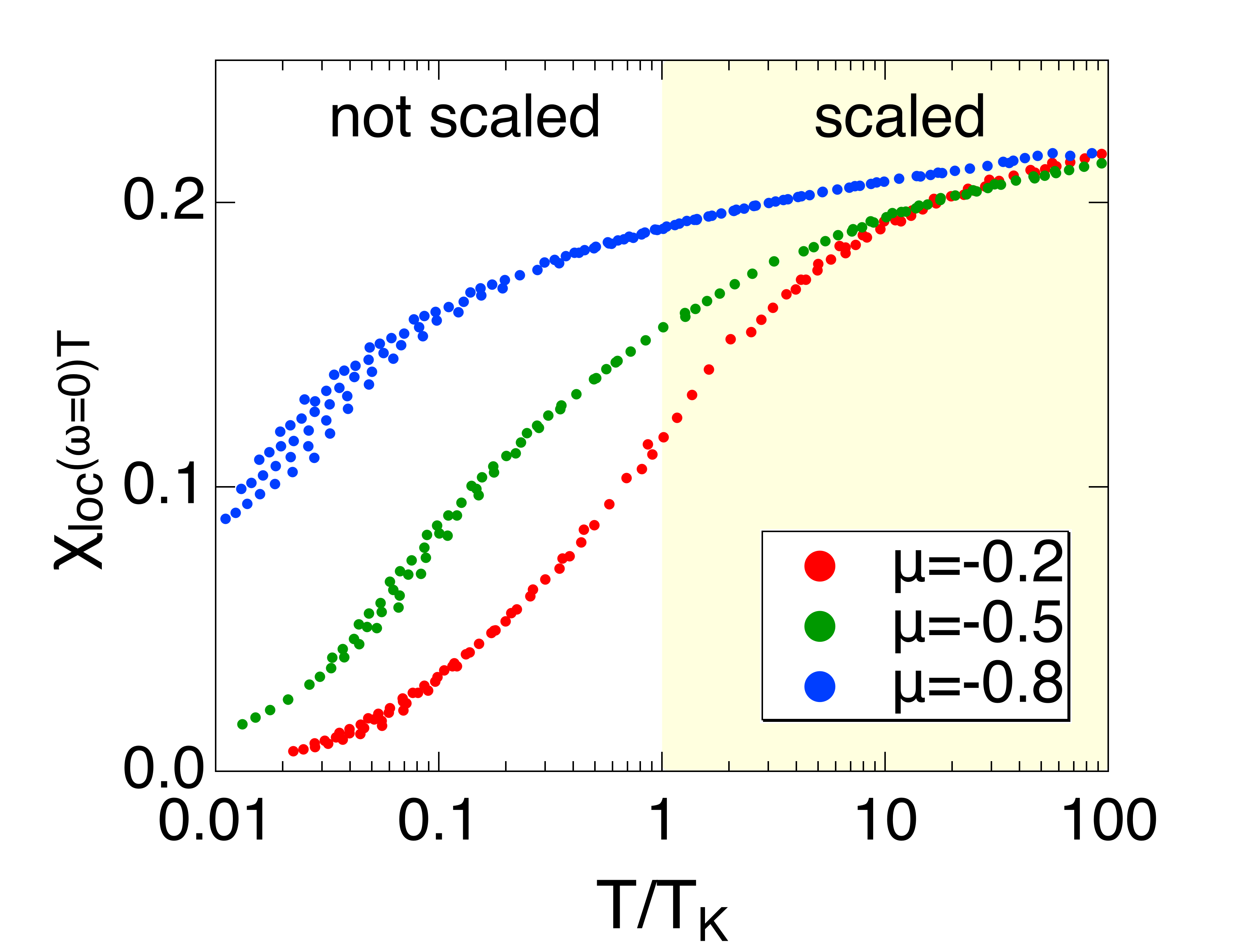}}
    \noindent\subfloat[]{\includegraphics[width=.24\textwidth]{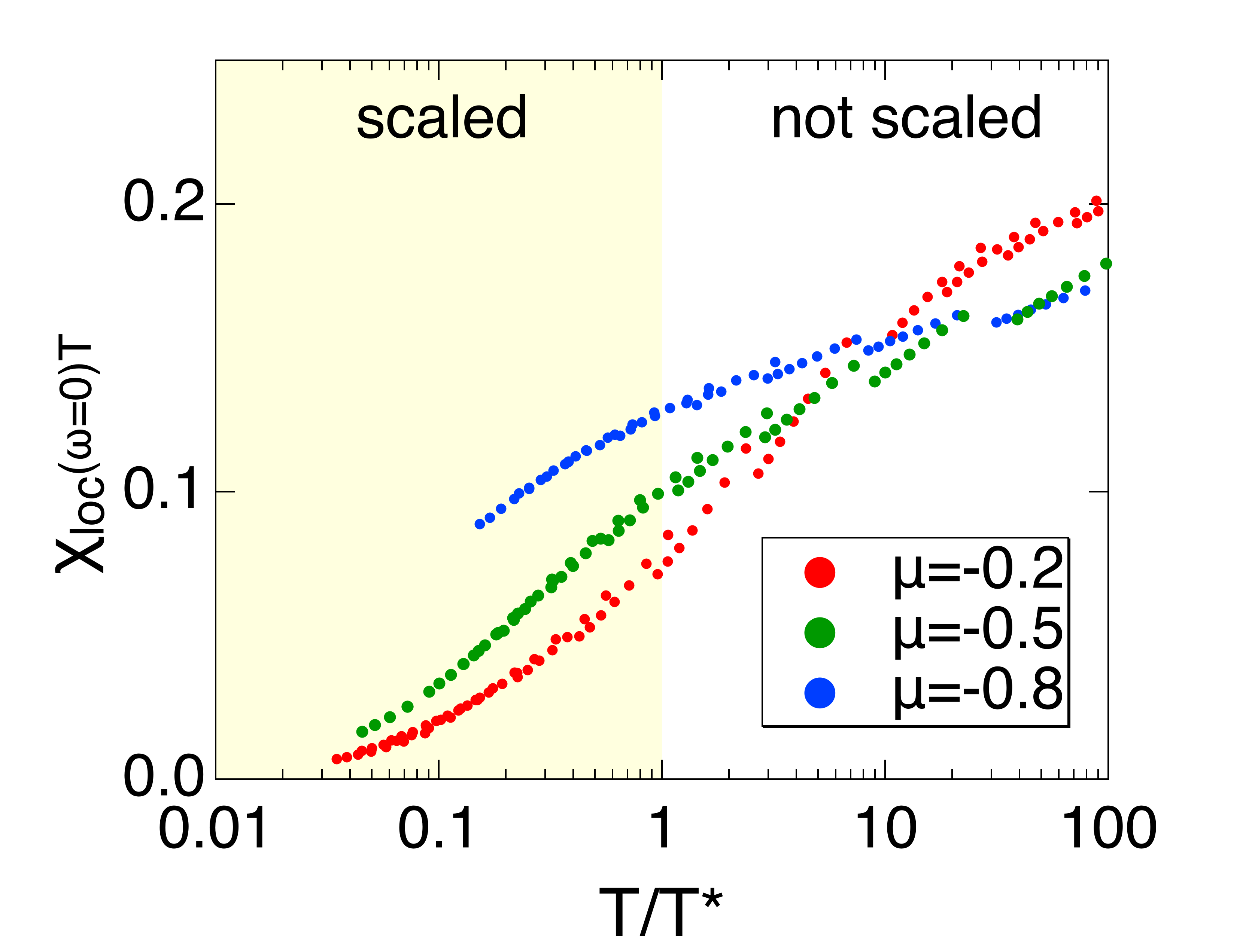}}
    \newline
    \subfloat[]{\includegraphics[width=.24\textwidth]{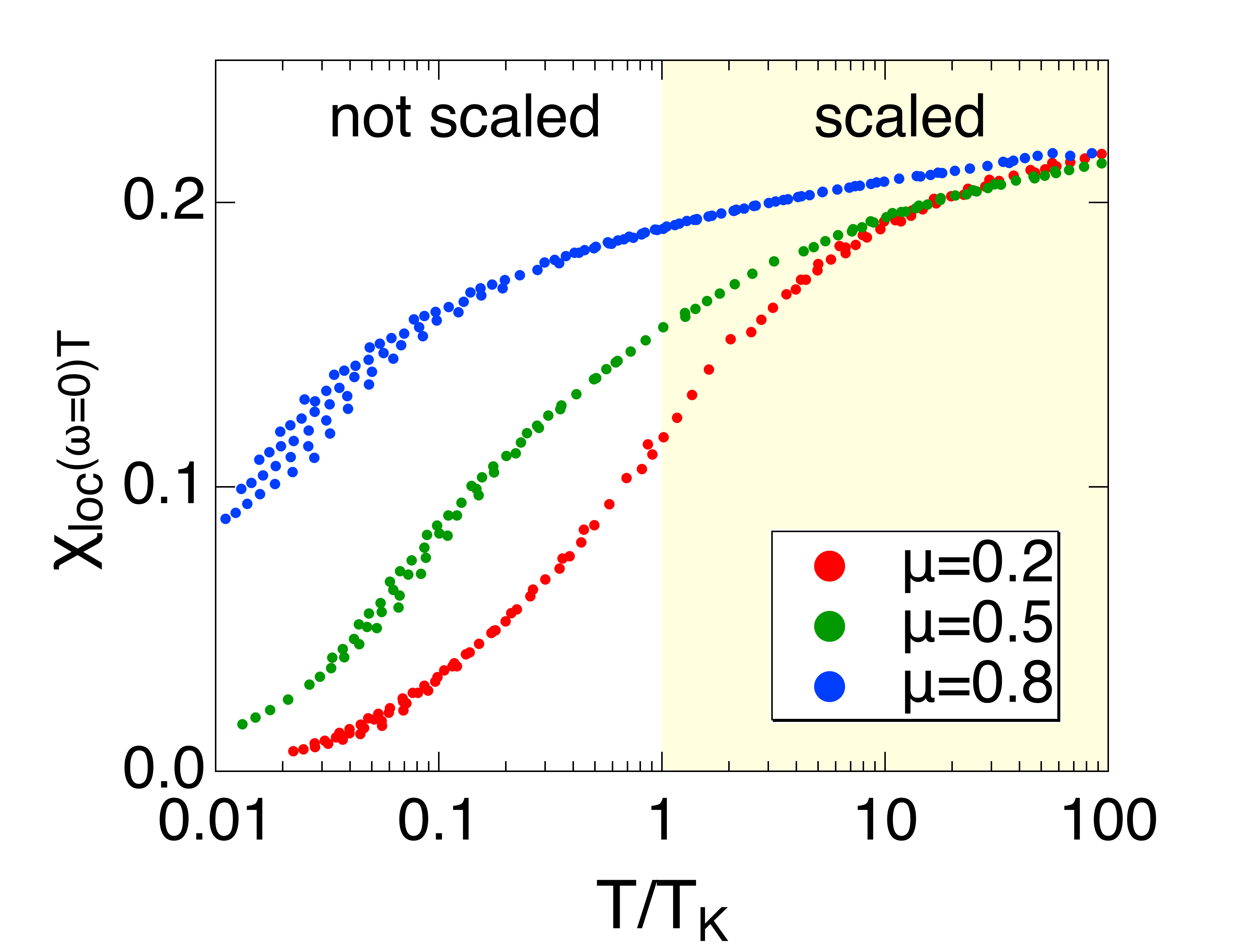}}
    \noindent\subfloat[]{\includegraphics[width=.24\textwidth]{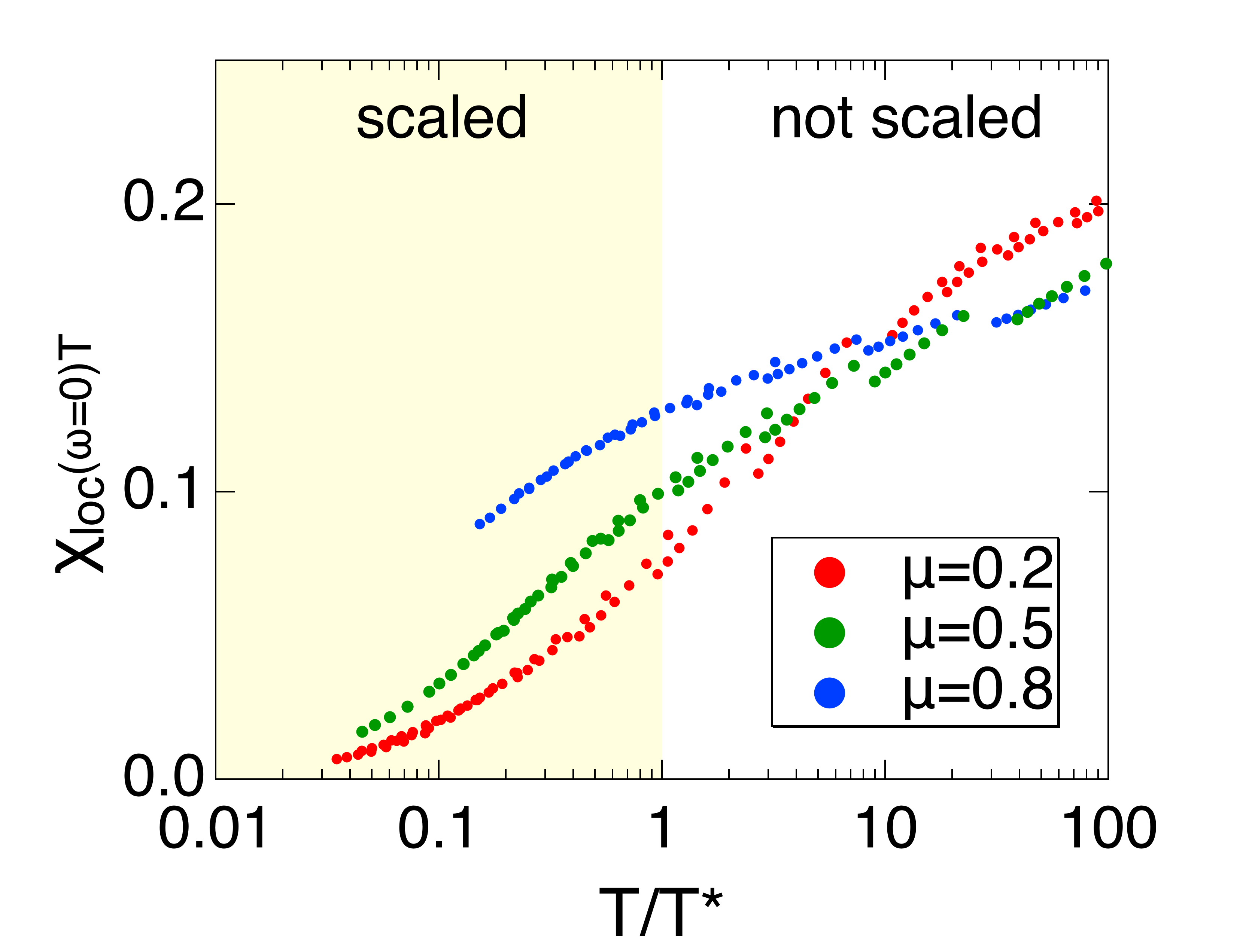}}
    \caption{Local static spin susceptibility scaled by $T_K$ (a) 
        and $T^*$ (b) of the Anderson lattice model with
        chemical potential of $\mu=-0.2,-0.5,-0.8$, 
        and scaled by $T_K$ (c) and $T^*$ (d) of the Anderson lattice model with 
        chemical potential of $\mu=0.2,0.5,0.8$.
        }
\end{figure}

The particle-hole symmetry of the undoped ($\mu=0.0$) is presented in Fig. 2. 
Figure 2 shows that the same temperature scales can be used to scale the hole-doped ($\mu>0.0$) cases of 
Fig. 2 (c) and (d) with the electron-doped ($\mu<0.0$) cases of Fig. 2 (a) and (b) 
as expected from the particle-hole symmetry of the undoped ($\mu=0.0$) model. 


When total number of electrons ($n=n_{f}+n_{c}$) is fixed, chemical potential $\mu$ is changed
as the temperature and hybridization vary. 
To confirm that the coherence temperature $T^*$ also scales the physical observables, 
we calculated physical observables when the total number of electrons $n$ is fixed to 
$1.078$, which corresponds to the high-$T$ limit of the $mu-0.8$ case in the main text. 

\begin{figure}[htbp]\label{fixedn}
    \centering
    \includegraphics[width=.4\textwidth]{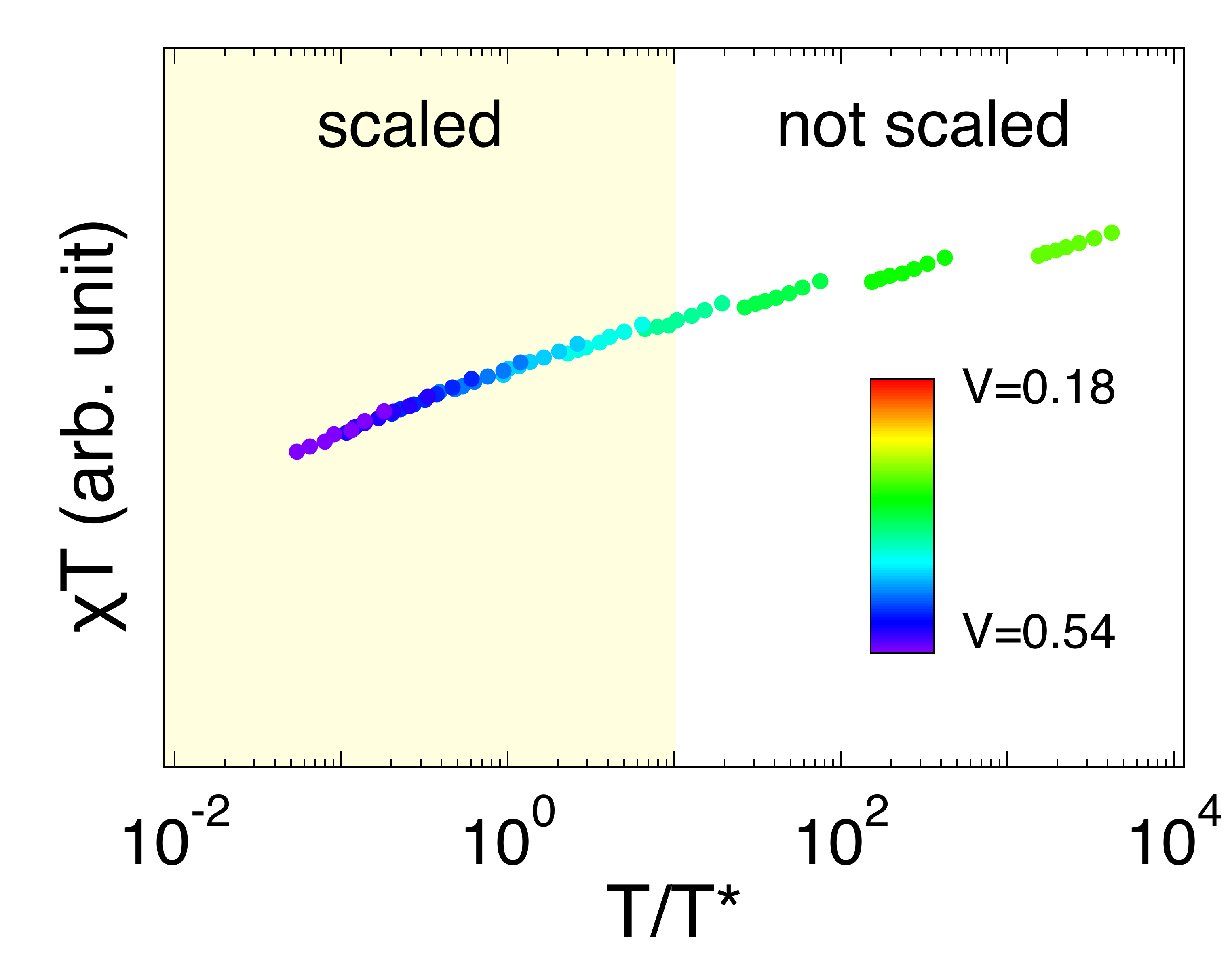}
    \caption{Local static spin susceptibility scaled by $T^*$ 
    with fixed $n=1.078$. }
\end{figure}

Fig. 3 shows how the local static spin susceptibility is scaled by 
the coherence temperature $T^*$ defined by proper effective coupling constant $J^{\text{latt}}$ 
which is obtained by the calculation with fixed chemical potential in the main text. 
It shows that $T^*$ scales the strong-coupling regime of the Anderson lattice model 
with fixed $n$.

%% file: author_list.tex
%

%
\author{Hanhim Kang} \affiliation{Department of Chemistry, Pohang University of Science and Technology, Pohang 790-784, Korea}
\author{Kristjan Haule} 
\affiliation{Department of Physics and Astronomy, Rutgers University, NJ, USA}
\author{Gabriel Kotliar}
\affiliation{Department of Physics and Astronomy, Rutgers University, NJ, USA}
\author{Piers Coleman}
\affiliation{Department of Physics and Astronomy, Rutgers University, NJ, USA}
\author{Ji-Hoon Shim$^\ast$}
\affiliation{Department of Chemistry, Pohang University of Science and Technology, Pohang 790-784, Korea}
\affiliation{Department of Physics, Pohang University of Science and Technology, Pohang 790-784, Korea}

%
%
%
\vskip 0.25cm